\documentclass[aip, sd, amsmath,amssymb,reprint]{revtex4-1}
\usepackage{graphicx}
\usepackage{dcolumn}
\usepackage{bm}
\usepackage{graphicx}
\usepackage{amsmath}
\usepackage{mathptm} 
\usepackage{hyperref}
\usepackage{xcolor}
\preprint{APS/123-QED}
\usepackage[normalem]{ulem}

\begin{document}
\title {Explosive Transitions in Complex Networks with Adaptive Competing Interactions}
\author{Umesh Kumar Verma}%
\email{umeshvermaphy@gmail.com}
\affiliation{Department of Physics, Babu Banarasi Das Northern India Institute of Technology, Lucknow UP-226028, India}

\begin{abstract}
Adaptation plays a central role in regulating collective behavior in complex systems. We study the collective dynamics of non-identical Stuart-Landau oscillators coupled through adaptive attractive-repulsive interactions. Without adaptation, oscillators coupled with only attractive coupling exhibit a continuous transition to synchronization. However, incorporating adaptive coupling, where the interaction strength evolves based on the global state of the system, induces an explosive transition to synchronization. When both attractive and repulsive couplings are present without adaptation, the system displays a continuous transition to synchronization and an abrupt transition to oscillation death. Remarkably, when adaptation is incorporated into this competing coupling framework, the system again exhibits an abrupt transition to synchronization. Interestingly, oscillation death occurs only in the absence of adaptation and is suppressed when adaptive coupling is present. These results are robust across different network topologies, including global, nonlocal, and scale-free networks, underscoring the versatility of adaptive mechanisms in controlling and stabilizing emergent dynamics in complex networks.

\end{abstract}
\maketitle




\section{Introduction}

The study of coupled dynamical systems offers a comprehensive framework for exploring the emergence of complex collective behavior in a wide range of real-world systems, including physical, biological, ecological, and social networks~\cite{watts,Pikovsky01}. In this framework, each unit is represented as a node, and the interactions between them are described as links, forming a networked structure that governs the overall dynamics. The interaction between nodes can be attractive, repulsive, or a combination of both, depending on the nature of the underlying system~\cite{Hong, Sathiyadevi,Chen,Nannan,Hens,Majhi,Dixit1}. Attractive and repulsive interactions are frequently observed in complex systems, and recent studies have highlighted their significance in nonlinear dynamics. For example, in neuronal networks, various firing patterns are induced through excitatory (attractive) and inhibitory (repulsive) synapses~\cite{Vogels,Myung}. Gene regulatory networks exhibit tunable motifs formed through the interlinking of positive and negative feedback loops, enabling robust control of gene expression and contributing to diverse dynamical regimes~\cite{Tian}. Attractive coupling tends to synchronize the dynamics of connected nodes, whereas repulsive coupling promotes desynchronization and suppression of oscillation. The interplay between these two types of coupling also gives rise to a wide range of emergent phenomena, including synchronization and suppression of oscillations.

Synchronization is one of the most fundamental collective phenomena, in which the dynamics of individual oscillators become aligned over time. It has been extensively studied in both theoretical models and real-world systems, including the rhythmic flashing of fireflies~\cite{Buck}, coordinated flocking of birds~\cite{Attanasi}, power grid dynamics~\cite{Buldyrev}, and neural activity in the brain~\cite{Adhikari}. The transition from an incoherent state to a synchronized state can occur either continuously or abruptly, depending on the system parameters and coupling mechanisms. An abrupt transition of this kind is termed explosive synchronization (ES)~\cite{Pazo}. Several mechanisms have been identified as key drivers of explosive synchronization (ES), including degree–frequency correlations in oscillator networks~\cite{Gardenes}, bimodal distributions of intrinsic frequencies~\cite{Li}, the inclusion of inertia in the oscillator dynamics~\cite{Tanaka}, and, more recently, attractive–repulsive interactions in multilayer networks~\cite{Verma22}.

Oscillation suppression refers to the cessation of oscillatory activity due to coupling, and it is broadly classified into two distinct types: amplitude death (AD)~\cite{prasad_review} and oscillation death (OD)~\cite{koseska_review}, distinguished by their origin and manifestation. AD corresponds to the stabilization of a pre-existing trivial steady state of the uncoupled system through coupling~\cite{prasad_review,strogatz_ad1990}. In contrast, oscillation death results from the emergence of coupling-induced fixed points, where the system stabilizes to either a symmetric homogeneous steady state (HSS) or a symmetry-breaking inhomogeneous steady state (IHSS)~\cite{koseska2013,banerjee_pre2014,Verma19}. For example, in laser systems, controlling unwanted fluctuations by suppressing oscillatory mechanisms results in more reliable outputs~\cite{Wei2007}. In the context of neuronal systems, suppressing irregular oscillations helps stabilize neural activity and provides insights into the treatment of disorders like epilepsy~\cite{Ermentrout1990}. The transition from an oscillatory state to a death state can occur via two distinct routes: a smooth, continuous (second-order) transition or a sudden, discontinuous (first-order) transition. Abrupt transitions to a death state, known as explosive death (ED), have been extensively studied~\cite{Bi,Verma,Verma1}. This phenomenon is characterized by the emergence of hysteresis or bistability, where oscillatory and death states coexist over a range of parameter values. Such bistable behavior has been observed in various physical~\cite{Herrero} and chemical systems~\cite{Bar-Eli}.

Traditionally, global synchronization has been understood as a bottom-up process, where local clusters of synchronized oscillators first form and then progressively merge into a globally coherent state as the strength of their coupling increases. However, if this hierarchical buildup is disrupted due to structural asymmetries, competitive couplings, or any perturbations, the system may remain fragmented. But near a critical threshold, these disconnected clusters can suddenly merge, leading to an abrupt and explosive transition to synchrony. In this study, we explore how abrupt transitions in complex networks can be induced and controlled through adaptive coupling with competing interactions. We focus on a feedback-driven mechanism in which the coupling strength is dynamically adjusted based on the collective state of the system. Specifically, the coupling evolves as a function of the global order parameter, allowing it to either strengthen or weaken depending on the number of synchronized oscillators. This adaptive strategy reflects a fundamental principle observed in many real-world systems. For instance, in the brain, neural connectivity is inherently dynamic synaptic strengths evolve in response to correlated activity among neurons~\cite{Do}. Learning is an adaptive process in which neural circuitry reorganizes itself based on experience and environmental feedback, allowing the brain to acquire and retain new information~\cite{Sawicki}. A similar principle can be observed in musical ensembles, where performers continuously adjust their timing and expression in response to the conductor’s guidance and auditory feedback from fellow musicians, leading to coordinated and harmonious performance. Unlike local adaptive schemes such as Hebbian learning~\cite{Avalos}, where coupling evolution depends on pairwise correlations, the present global-order-parameter–based adaptation captures system-wide feedback, enabling the regulation of collective coherence through macroscopic dynamics.

The form of adaptation we utilize, governed by the global order parameter, was first proposed by Filatrella et al.~\cite{Filatrella} in the context of Kuramoto oscillator networks. This approach was later generalized and analytically examined by Zou and Wang~\cite{Zou}. Recently, Zhang et al.~\cite{Zhang} demonstrated that adaptive coupling mediated by the global order parameter can induce abrupt synchronization transitions in phase oscillator networks. Verma~\cite{Verma25} further showed that adaptive feedback-type coupling leads to explosive synchronization transitions accompanied by hysteresis in amplitude oscillator systems. However, these studies primarily considered specific coupling schemes and did not address the effects of competing attractive and repulsive adaptive interactions or their influence on oscillation-death states. Motivated by this gap, we examine the effects of adaptive attractive–repulsive coupling on collective dynamics in Stuart–Landau oscillator networks. We consider an ensemble of non-identical Stuart–Landau (SL) oscillators coupled through adaptive attractive–repulsive interactions. We begin by analyzing the case with purely attractive coupling. Without adaptation, the system exhibits a continuous (second-order) transition to synchronization. However, introducing adaptive coupling leads to a discontinuous (first-order) transition to the synchronized state. When both attractive and repulsive couplings are present, the system shows a second-order transition to synchronization and a first-order transition to oscillation death in the absence of adaptation. Remarkably, incorporating adaptive coupling not only triggers a first-order transition to synchronization but also completely suppresses the transition to oscillation death. Thus, oscillation death arises only under competing interactions without adaptation and disappears once adaptation is included. These results are robust across different network topologies, including global, nonlocal, and scale-free networks.

\section{Model}
We consider a system of $N$ non-identical Stuart-Landau (SL) oscillators that are coupled with adaptive attractive-repulsive coupling. The dynamics of each coupled SL oscillator can be represented by the following equation:
\begin{eqnarray}
\dot{x}_{i}&=&P_{i}x_{i}-\omega_i y_{i} +\frac{\lambda_{a}R^{\alpha}}{N}\sum_{j=1}^{N}(x_{j}-x_{i}),\nonumber\\
\dot{y}_{i}&=&P_{i}y_{i}+\omega_i x_{i}-\frac{\lambda_{r}R^{\alpha}}{N}\sum_{j=1}^{N}(y_{j}-y_{i})
\label{eq1}  		
\end{eqnarray}
 Here, $P_i = \rho_i - x_i^2 - y_i^2$ for $i = 1,2,\ldots,N$, where $x_i$ and $y_i$ are the state variables of the $i$th Stuart--Landau (SL) oscillator representing its position in phase space. The intrinsic frequency $\omega_i$ of the $i$th oscillator is drawn randomly from a uniform distribution, $\omega_i \in [1,2]$. The parameter $\rho_i$ characterizes the intrinsic amplitude of the $i$th oscillator and governs its local dynamics. Unless otherwise stated, we set $\rho_i = 1$. The parameters $\lambda_a$ and $\lambda_r$ denote the strengths of attractive and repulsive couplings, respectively. The adaptation exponent $\alpha$ controls the degree of adaptation, for $\alpha = 0$ corresponds to the non-adaptive case, while $\alpha = 1$ represents maximal adaptation. Unless otherwise stated, the system size is fixed at $N = 200$.

{\em \underline{Order parameters:}} The synchronization order parameter $R$ is defined to quantify the collective phase coherence of the coupled systems and is given by 
\begin{equation}
    Re^{\iota \psi}=\frac{1}{N}\sum_{j=1}^{N}e^{\iota\theta_j}
\end{equation}
where $j=1,2,3....,N$. $\psi$ is the average phase of the oscillators. The instantaneous phase of each oscillator $\theta_i$ is calculated as $\theta_i=\tan^{-1}(y_i/x_i)$. The value of the order parameter $R$ lies between 0 and 1. For coherence state value of $R\approx 1$, while for incoherence state $R \approx 0$.

To quantify the collective amplitude of the coupled systems, we define the amplitude order parameter $A$ as
\begin{equation}
A(\lambda_l) = \frac{1}{N}\sum_{i=1}^N \left[\langle x_{i,max} \rangle_t - \langle x_{i,min}\rangle_t \right].\nonumber
\end{equation}

where $x_{i,max}$ and $x_{i,min}$ represent the time-averaged maximum and minimum values of the $i^{th}$ oscillator, respectively. $A(\lambda_l)=0$, when all oscillators are in the death state, whereas when all oscillators are in the oscillatory state, $A(\lambda_l)>0$. Here $\lambda_l$ is $\lambda_a$ or $\lambda_r$. Both the order parameter $R$ and $A$ are calculated in both forward and backward directions. In the forward continuation $R$ or $A$ is  calculated at $\lambda_l = 0$, for random initial conditions and \(\lambda_l\) is then increased from \(\lambda_l(0)\) to \(\lambda_l(\max)\) in increments of \(\delta\lambda_l = 0.02\). For each step, the final state obtained at the preceding value of \(\lambda_l\) is used as the initial condition for the next one. The backward sweep follows the same procedure, but with \(\lambda_l\) decreased from \(\lambda_l(\max)\) to \(\lambda_l(0)\). All numerical simulations of Eq.~\ref{eq1} are performed using the fourth-order Runge-Kutta method, and the initial conditions are sampled randomly from the interval \([-1, 1]\).

\begin{figure}
\centering
\includegraphics[width=0.48\textwidth]{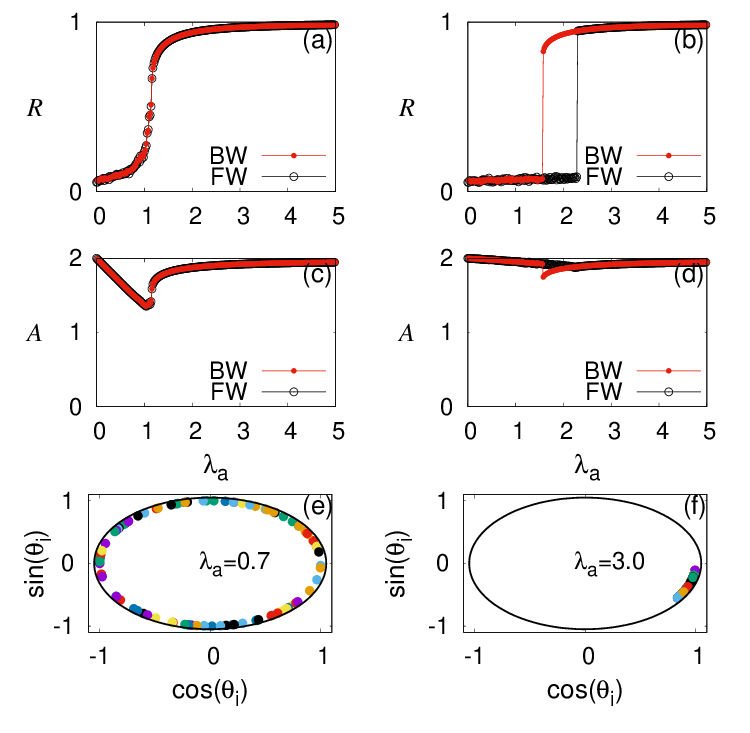}
\caption{The synchronization order parameter $R$ and amplitude order parameter $A$ are computed for attractively coupled Stuart--Landau oscillators as functions of the attractive coupling strength
$\lambda_a$, using both forward continuation (black curves) and backward continuation (red curves). Panels (a,c) correspond to the non-adaptive case ($\alpha = 0$), while panels (b,d) show the adaptive case ($\alpha =1$). Panels (e,f) display representative time series of oscillators for the adaptive case ($\alpha = 1$) at (e) $\lambda_a = 0.7$ (incoherent state) and (f) $\lambda_a = 3.0$ (synchronized state). Panels (g,h) show the corresponding phase portraits in the $(\cos\theta_i,\sin\theta_i)$ plane for $\alpha = 1$ at (g) $\lambda_a = 0.7$ and (h) $\lambda_a = 3.0$. The repulsive coupling strength is fixed at $\lambda_r = 0.0$}
\label{fig1}
\end{figure}

\section{Global coupling  topology}

{\em \underline{Results for attractive coupling:}} We first consider a case where SL oscillators are only coupled with attractive coupling (i.e. $\lambda_r=0$). In this scenario, we initially analyze the dynamics of the coupled SL oscillators without adaptation (i.e. $\alpha=0$). The synchronization and amplitude order parameters are calculated and presented in Fig.~\ref{fig1}(a,c), respectively. In Fig.~\ref{fig1}(a), synchronization order parameter $R$ shows that the transition from the desynchronized state to the synchronized state is continuous in both forward and backward directions. In Fig.~\ref{fig1}(c), the amplitude order parameter $A$ indicates that the amplitude of the coupled system gradually decreases during the transition and then recovers to its original value once synchronization is achieved. Notably, we observe that no transition from oscillatory state to oscillation death state occurs in the case of attractively coupled SL oscillators.
Further, we consider a case where SL oscillators coupled with adaptive attractive coupling (i.e. $\alpha=1$). In this case, we observe a sudden transition in the synchronization order parameter in both forward and backward directions, which is shown in Fig.~\ref{fig1}(b). This result indicates that introducing adaptation into the system induces an explosive (first-order) transition to synchronization in the coupled oscillators. The amplitude order parameter $A$ for $\alpha=1$ is plotted in Fig.~\ref{fig1}(d), which indicates that the amplitude is mostly the same in this process. The time series plot of the coupled SL oscillators before and after transitions for $\alpha=1$, is shown in Fig.~\ref{fig1}(e,f) respectively. We have also plotted the phase portrait plot $(\cos\theta_i, \sin\theta_i)$ of coupled SL oscillators before and after transition at time $(t=10)$, which are shown in Fig.~\ref{fig1}(g,h) respectively.

\begin{figure}
    \centering
    \includegraphics[width=0.48\textwidth]{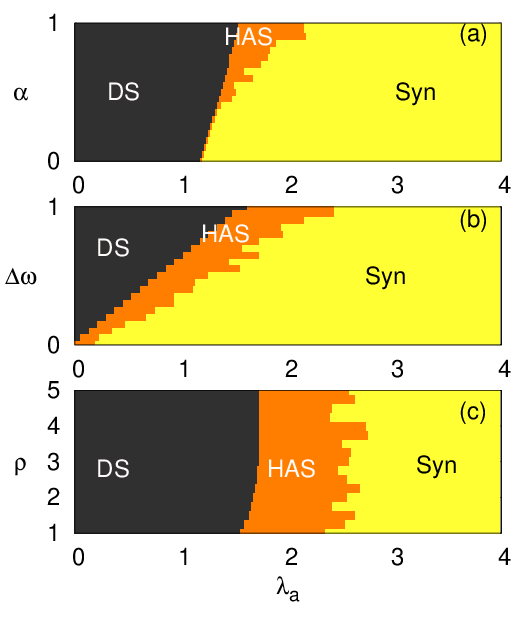}
    \caption{Phase diagrams of Stuart--Landau (SL) oscillators coupled via attractive interaction in different parameter spaces: (a) $(\lambda_a-\alpha)$,  (b) $(\lambda_a-\Delta\omega)$, and (c) $(\lambda_a-\rho)$. Here, $\alpha$ is the adaptation parameter, $\Delta\omega$ denotes the width of the intrinsic frequency distribution. The regions labeled DS, Syn, and HAS correspond to the desynchronized state, synchronized state, and the hysteresis area within the synchronized regime, respectively. The repulsive coupling strength is fixed at $\lambda_r = 0.0$. }
    \label{fig2}
\end{figure}

The phase diagram $(\lambda_a-\alpha)$ of attractively coupled SL oscillators is shown in Fig.~\ref{fig2}(a). Here DS, Syn, and HAS represent the desynchronized state, synchronized state, and hysteresis area in the synchronized regime.  This figure also confirms that for $\alpha=0$, the coupled oscillators do not show hysteresis where both synchronized and desynchronized oscillation co-exist. By increasing the adaptation strength $\alpha$ increases, the width of the hysteresis region expands noticeably in the $(\lambda_a,\alpha)$ parameter space. Stronger adaptation amplifies the positive feedback between the collective dynamics and the coupling strength, thereby stabilizing the synchronized state over a wider range of coupling values. Further, we have examined the role of the width of the frequency distribution. For this, we calculate the phase diagram $(\lambda_a-\Delta\omega)$ for $\alpha = 1$, where frequencies of the oscillators are
distributed as $\omega_i = \omega_o + \Delta\omega$. Where $\omega_o=1$ and $\Delta\omega$ is the width of the uniform distribution. The phase diagram $(\lambda_a-\Delta\omega)$ is shown in Fig.~\ref{fig2}(b), which indicates that the transition points occur at a higher value of $\lambda_a$ as we increase the distribution width. The phase diagram in the $(\lambda_a-\rho)$ parameter space is shown in Fig.~\ref{fig2}(c). 
It is observed that increasing the parameter $\rho$ shifts both the forward and backward transition points toward higher values of $\lambda_a$, indicating that stronger intrinsic amplitude requires larger coupling strength to achieve synchronization.

\begin{figure}
    \centering
    \includegraphics[width=0.48\textwidth]{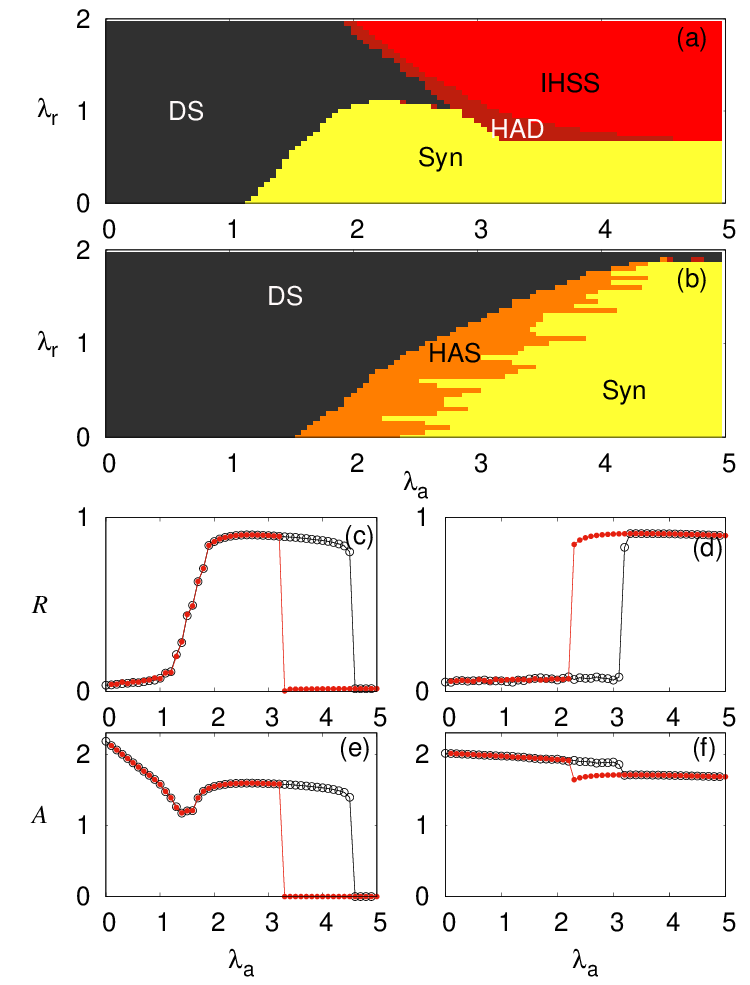}
    \caption{Phase diagram of SL oscillators coupled with attractive and repulsive coupling in the parameter space: (a) $(\lambda_a-\lambda_r)$ for $\alpha=0$, and (b) $(\lambda_a-\lambda_r)$ for $\alpha=1$. The synchronization order parameter $R$ and amplitude order parameter $A$ are calculated in both forward and backward directions: (c,e) for $\lambda=0.7$. and $\alpha=0$, and (d,f) for $\lambda=0.7$. and $\alpha=1$.}
    \label{fig3}
\end{figure}

{\em \underline{Results for attractive-repulsive coupling:}} Next, we examine the case where SL oscillators are coupled via attractive-repulsive coupling. We begin by analyzing the system without adaptation by setting $\alpha = 0$ and plotting the phase diagram in the $(\lambda_a-\lambda_r)$ parameter space, as shown in Fig.~\ref{fig3}(a). For repulsive coupling strength $\lambda_r < 1.1$, the coupled system exhibits synchronized oscillations. However, as $\lambda_r$ exceeds this threshold ($\lambda_r > 1.1$), the synchronized state becomes unstable, and a transition from the oscillatory state to an oscillation death state is observed, characterized by the emergence of an inhomogeneous steady state (IHSS). Notably, this transition is accompanied by hysteresis, indicating a bistable region where both oscillatory and death states coexist. The synchronization order parameter $R$ and amplitude order parameter $A$ are plotted in Fig.~\ref{fig3}(c,e) for $\alpha=0$, and $\lambda_r=0.7$, which indicates that a second-order transition from a desynchronized state to a synchronized state and a first-order transition from an oscillatory state to a death state with hysteresis. 

We then investigate the effect of adaptation by plotting the phase diagram in the parameter space $(\lambda_a-\lambda_r)$  for $\alpha = 1$, as shown in Fig.~\ref{fig3}(b). This diagram reveals that, in the presence of adaptation, the system exhibits a transition from a desynchronized state to a synchronized state, also accompanied by hysteresis. However, unlike the non-adaptive case, no transition to the oscillation death state is observed.  This suggests that the incorporation of adaptation effectively sustains oscillatory dynamics and inhibits the onset of oscillation death. The order parameter $R$ and $A$ plotted in Fig.~\ref{fig3}(d,f) respectively.  These plots confirm a first-order transition from a desynchronized to a synchronized state, and the absence of any transition to an oscillation death state.

\section{ Non-local coupling topology}
To check the generality of our results, we now consider a network of $N$ Stuart–Landau oscillators coupled through a nonlocal attractive–repulsive coupling topology. In this configuration, each oscillator interacts with a finite number of neighbors on either side. The dynamics of each coupled SL oscillator can be represented by the following equation:
\begin{eqnarray}
\dot{x}_{i}&=&P_{i}x_{i}-\omega_i y_{i} +\frac{\lambda_{a}R^{\alpha}}{2P}\sum_{j=i-P}^{i+P}(x_{j}-x_{i}),\nonumber\\
\dot{y}_{i}&=&P_{i}y_{i}+\omega_i x_{i}-\frac{\lambda_{r}R^{\alpha}}{2P}\sum_{j=i-P}^{i+P}(y_{j}-y_{i})
\label{eq2}  		
\end{eqnarray}

Where $P_{i} = \rho_i - x_{i}^2 - y_{i}^2$ with $i = 1, 2, 3, \dots, N$. Each node $i$ is connected to $P$ nearest neighbors in both directions along a ring topology, with the coupling range $P \in [1, N/2]$. When $P = 1$, the network corresponds to \textit{local coupling}, where each node interacts only with its immediate neighbors. When $P = N/2$, the network exhibits \textit{global coupling}, meaning that each node interacts with all other nodes in the system. For intermediate values of $P$ (i.e., $1 < P < N/2$), the network demonstrates \textit{non-local coupling }. This parameter $P$ represents the range of interaction, with the coupling radius defined as $CR = P/N$.

\begin{figure}
\centering
\includegraphics[width=0.48\textwidth]{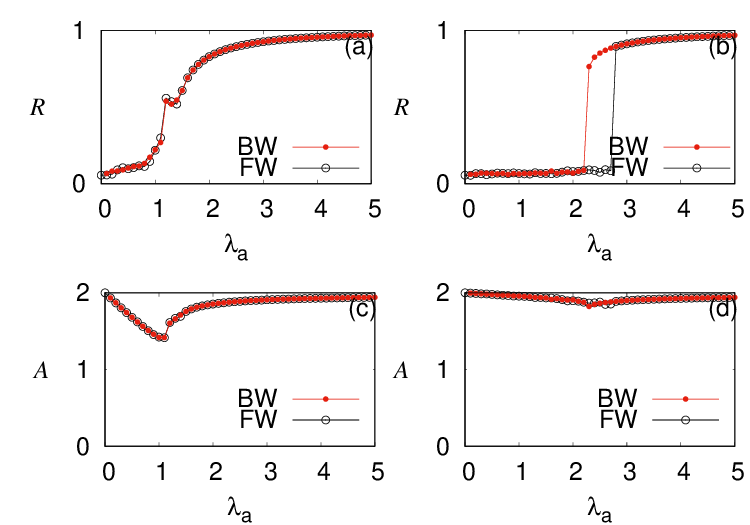}
\caption{The order parameters $R$ and $A$ are computed for SL oscillators coupled through a nonlocal attractive–repulsive coupling in both forward and backward directions. (a,c) $\alpha=0$ (without adaptation), (b,d) $\alpha=1$ (with adaptation).  The other parameters are $CR=0.1$, $\lambda_r=0.0$, $\rho=1$, and $N=200$.}
\label{fig4}
\end{figure}

\begin{figure}
\centering
\includegraphics[width=0.48\textwidth]{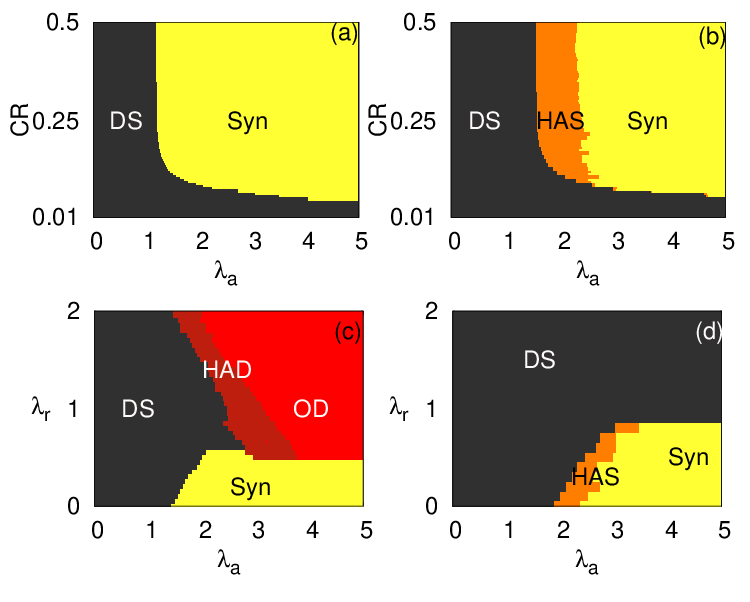}
\caption{Phase diagrams of SL oscillators. Panels (a,b) show the phase diagram in the $(\lambda_a-\mathrm{CR})$ plane for
attractively coupled SL oscillators with (a) $\alpha = 0$ (non-adaptive case) and (b) $\alpha = 1$ (adaptive case), where $\mathrm{CR}$ denotes the coupling radius. Panels (c,d) show the phase diagram in the $(\lambda_a-\lambda_r)$ plane for SL oscillators with competing attractive--repulsive coupling at a fixed coupling radius $\mathrm{CR} = 0.125$, with (c) $\alpha = 0$ and (d) $\alpha = 1$.}
\label{fig5}
\end{figure}

{\em \underline{Results for attractive coupling:}}

Next, we consider SL oscillators coupled with non-local attractive coupling and compute the order parameters $R$ and $A$ for a fixed coupling radius $CR = 0.1$, and $a = 0$, as shown in Fig.~\ref{fig4}(a,c), respectively. Figure~\ref{fig4}(a) illustrates a continuous transition from a desynchronized state to a synchronized state as the attractive coupling strength increases. In Fig.~\ref{fig4}(b), the order parameter $A$ shows that the amplitude of the coupled system initially decreases. However, once synchronization is achieved, the system regains and maintains its original amplitude.

We further examine the effect of adaptation by setting $a = 1$ for the same coupling radius $CR = 0.1$. The synchronization order parameter $R$, shown in Fig.~\ref{fig4}(c), exhibits a sudden transition in both forward and backward directions. These transitions occur at different values of $\lambda_a$, indicating hysteresis in the system. Fig.~\ref{fig4}(d) reveals that the amplitude order parameter $A$ does not display any transition from oscillatory states to death states. 

To examine the influence of coupling radius, we construct the parameter space for attractively coupled oscillators. The phase diagram in the $(\lambda_a-CR)$ plane, corresponding to $a = 0$ and $\lambda_r = 0$, is presented in Fig.~\ref{fig5}(a). The diagram reveals that for a small coupling radius, or for a lower number of connections, the system remains in the desynchronized state without exhibiting a transition to synchronization. As the coupling radius increases, a clear transition from the desynchronized state to the synchronized state is observed. We also plot the parameter space $(\lambda_a-CR)$, for $a=1$, which is presented in Fig.~\ref{fig5}(b). This diagram similarly indicates that for a small coupling radius, the system remains in desynchronized state; however, as the coupling radius increases, the system undergoes a transition from the desynchronized to the synchronized state, accompanied by hysteresis. These results demonstrate that incorporating adaptation in the system induces bistability near the transition point.

\begin{figure}
\centering
\includegraphics[width=0.48\textwidth]{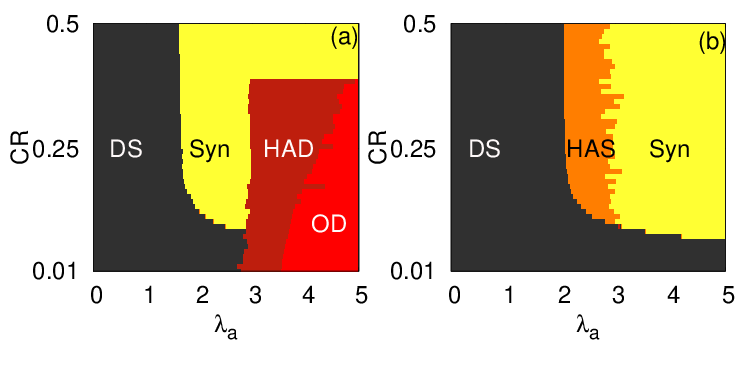}
\caption{The phase diagram $(\lambda_a-CR)$ of Stuart–Landau oscillators coupled with both attractive and repulsive coupling: (a) $\alpha=0$, (b) $\alpha=1$. Here $\lambda_r=0.50$.}
\label{fig6}
\end{figure}

{\em \underline{Results for attractive-repulsive coupling:}}
We further examine the dynamics of SL oscillators coupled via attractive–repulsive nonlocal coupling. In this case, we plot the parameter space $(\lambda_a - \lambda_r)$ for a coupling radius of $CR = 0.125$ (i.e., $P = 25$) and $a = 0$, as shown in Fig.~\ref{fig5}(c). The figure reveals that for a weaker repulsive coupling strength $\lambda_r$, the coupled system undergoes a transition to synchronization. However, as the repulsive coupling strength increases, the synchronized domain disappears, and a transition from the oscillatory state to the death state occurs, accompanied by hysteresis. We also plot the parameter space $(\lambda_a - \lambda_r)$ for $a = 1$, shown in Fig.~\ref{fig5}(d). This figure indicates that for lower values of $\lambda_r$, the coupled system displays a transition to synchronization with hysteresis. However, for $\lambda_r > 0.8$, the system shows only desynchronized oscillations.

Furthermore, Fig.~\ref{fig6} presents the phase diagram in the $(\lambda_a - CR)$ plane for a fixed repulsive coupling strength $\lambda_r = 0.5$. For $a = 0$, the system exhibits a desynchronized state and a transition from the oscillatory to the death state at lower values of the coupling radius $CR$. As $CR$ increases, transitions from desynchronized to synchronized states, as well as from oscillatory to death states, are observed as shown in Fig.~\ref{fig6}(a). When we set the adaptation parameter $a = 1$, the system shows a desynchronized state for a small coupling radius; however, with further increase in $CR$, a transition from the desynchronized to the synchronized state occurs, accompanied by hysteresis, as depicted in Fig.~\ref{fig6}(b).

\section{Scale-free coupling topology}
To further examine the influence of network architecture on emergent dynamics, we consider a network of $N$ SL oscillators coupled through a scale-free attractive–repulsive coupling topology. A scale-free network is a more realistic and heterogeneous topology found in many natural and technological systems. The dynamics of each coupled oscillator in this topology are governed by the following equation:

\begin{eqnarray}
\dot{x}_{i}&=&P_{i}x_{i}-\omega_i y_{i} +\frac{\lambda_{a}R^{\alpha}}{k_i}\sum_{j=1}^{N}A_{ij}(x_{j}-x_{i}),\nonumber\\
\dot{y}_{i}&=&P_{i}y_{i}+\omega_i x_{i}-\frac{\lambda_{r}R^{\alpha}}{k_i}\sum_{j=1}^{N}A_{ij}(y_{j}-y_{i})
\label{eq3}  		
\end{eqnarray}

Where $P_{i}=\rho_i-x_{i}^2-y_{i}^2$ and $A_{ij}$ is the adjacency matrix of the network. The degree of node $i$ is defined as $k_i=\sum_{i=1}^N A_{ij}$, representing the number of connections it has. The average degree of the network is defined as the $<k>=<k_i>$.

\begin{figure}
\centering
\includegraphics[width=0.48\textwidth]{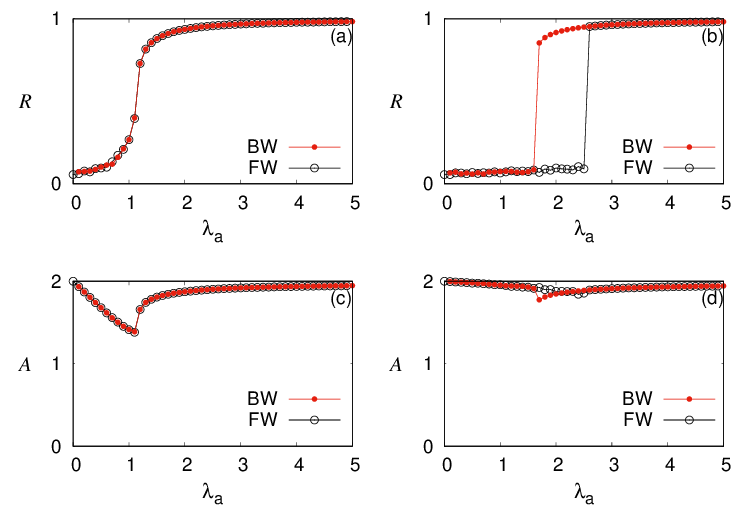}
\caption{The order parameters $R$ and $A$ are computed for SL oscillators coupled with attractive scale-free coupling in both forward and backward directions. (a,c) $\alpha=0$ (without adaptation), (b,d) $\alpha=1$ (with adaptation).  The other parameters are $<k>=40$, and $\lambda_r=0.0$.}
\label{fig7}
\end{figure}

{\em \underline{Results for attractive coupling:}}
We now consider the case where SL oscillators are coupled via an attractive scale-free network topology. The average degree of the network is fixed at $\langle k \rangle = 40$, and $\lambda_r = 0$. We compute the synchronization order parameter $R$ and the amplitude order parameter $A$ for both adaptive and non-adaptive networks. Figure~\ref{fig7}(a) shows a continuous transition from a desynchronized state to a synchronized state in the absence of adaptation. In contrast, when adaptation is introduced, we observe a first-order transition from the desynchronized state to the synchronized state, as depicted in Fig.~\ref{fig7}(b). Furthermore, Figs.~\ref{fig7}(c,d) indicate that no transition occurs from oscillatory dynamics to oscillation death, regardless of whether adaptation is present or absent.

\begin{figure}
\centering
\includegraphics[width=0.48\textwidth]{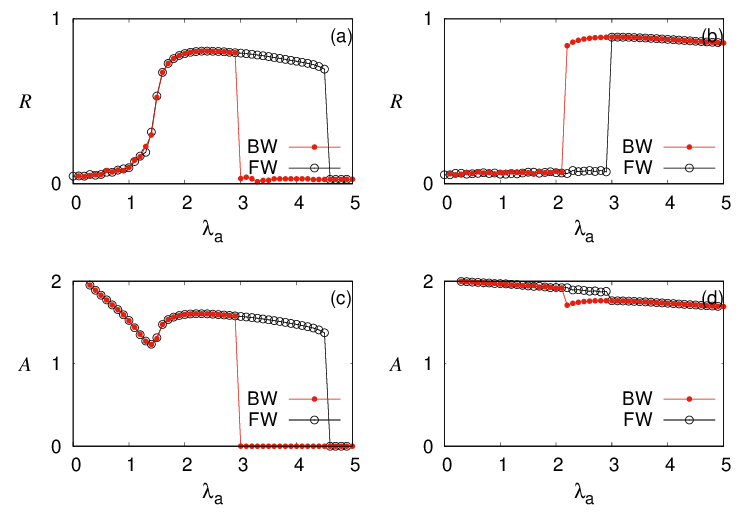}
\caption{The order parameters $R$ and $A$ are computed for attractive-repulsive scale-free coupled Stuart–Landau oscillators in both forward and backward directions. (a,c) $\alpha=0$ (without adaptation), (b,d) $\alpha=1$ (with adaptation).  The other parameters are $<k>=40$, $\lambda_r=0.5$.}
\label{fig8}
\end{figure}

{\em \underline{Results for attractive–repulsive coupling:}}  
Next, we investigate the dynamics of SL oscillators under an attractive–repulsive scale-free network topology. Here, we fix the average degree of the network at $\langle k \rangle = 40$, and set $\lambda_r = 0.5$. For this coupling scheme, we evaluate the order parameters $R$ and $A$ in both adaptive and non-adaptive cases. In the absence of adaptation, the system exhibits a continuous transition from desynchronized state to synchronized state, as shown in Fig.~\ref{fig8}(a), and a first-order transition from oscillatory dynamics to the oscillation death state, as shown in Fig.~\ref{fig8}(c). However, when adaptation is incorporated, the synchronization transition becomes abrupt, leading to a discontinuous jump to a synchronized state [see Fig.~\ref{fig8}(b)]. Moreover, in the adaptive attractive–repulsive scale-free network, no transition to oscillation death is observed, as shown in Fig.~\ref{fig8}(d), indicating that adaptation preserves or restores oscillatory behavior in scale-free network topologies.

\section{Conclusion}
In summary, we have investigated the dynamics of non-identical Stuart–Landau oscillators coupled through adaptive attractive-repulsive interactions. Our results demonstrate that the nature of the synchronization transition strongly depends on both the type of coupling and the presence of adaptation. Without adaptation, attractively coupled oscillators undergo a continuous (second-order) transition to synchronization, whereas the inclusion of adaptive coupling induces an abrupt (first-order or explosive) transition to synchronization. In the case of competing attractive–repulsive interactions, the system exhibits a continuous transition to synchronization and an explosive transition to oscillation death in the absence of adaptation. In contrast, when adaptation is incorporated into the competing coupling framework, the system displays a first-order transition to synchronization and no transition to oscillation death. Therefore, adaptation not only changes the synchronization transition from continuous to discontinuous but also fully suppresses oscillation death, enabling the network to maintain or recover oscillatory behavior. 

We have further extended our analysis to non-locally and scale-free coupled oscillators, confirming that the observed dynamical features persist beyond globally coupled networks. This demonstrates the broad robustness of the phenomena across diverse topologies and highlights the central role of adaptive mechanisms in shaping and stabilizing collective dynamics.The results of this work offer valuable insights into the dynamics of real-world complex systems such as the brain, where adaptation~\cite{Van} plays a critical role in maintaining function. More broadly, our work suggests that adaptive coupling can serve as an effective control strategy for tuning synchronization and preventing oscillation suppression in real-world systems ranging from neural circuits to power grids and gene regulatory networks.

\end{document}